\documentstyle[aps,prd,eqsecnum]{revtex}
\input psfig.sty

\begin{document}
\draft

\title{Comment on ``Ruling out chaos in compact binary systems''}

\author{Neil J. Cornish${}^\dagger$ and Janna Levin${}^*$}
\address{${}^\dagger$ Department of Physics, Montana State University, Bozeman, MT 59717}
\address{${}^* DAMTP$, Cambridge University,
Wilberforce Rd., Cambridge CB3 0WA }

\twocolumn[\hsize\textwidth\columnwidth\hsize\csname
           @twocolumnfalse\endcsname

\maketitle
\widetext

\begin{abstract}

In a recent {\it Letter}, Schnittman and Rasio \cite{sr}
argue that they have
ruled out chaos in compact binary systems since they find no 
positive Lyapunov exponents. 
In stark contrast, we find that the chaos discovered in the
original paper under discussion, J.Levin, PRL, {\bf 84} 3515 (2000),
is confirmed by the presence of
positive Lyapunov exponents.

\end{abstract}
\medskip
\noindent{04.30.Db,97.60.Lf,97.60.Jd,95.30.Sf,04.70.Bw,05.45}
\medskip
]

\narrowtext

\setcounter{section}{1}

{}
\vfill\eject
\newpage

Intuitively, black hole binaries are obvious candidates
for chaotic dynamics. Firstly, they are highly non-linear systems
and chaos is an expression of extreme non-linearity.  Secondly,
there are isolated unstable orbits around a Schwarzschild
black hole, and unstable orbits are a red flag for the onset
of chaos. Surely the orbits only become more unstable and more 
complex when there are two black holes.

There is no question that there is
chaos in black hole binaries for some range of parameters
\cite{{maeda},{mepn}}.
Yet the authors of Ref.\ \cite{sr} claim to rule out chaos by finding
no positive Lyapunov
exponents along the fractal of Ref.\ \cite{mepn}. 
Before carrying out a detailed calculation,
there is reason to be suspicious
of the absence of positive Lyapunov exponents.
Even the Schwarzschild solution
has positive Lyapunov exponents for unstable
circular orbits \cite{corn1}.
Though completely integrable
and so non-chaotic, the orbits are unstable.
{\it As long as there are unstable circular orbits in the PN equations,
there are positive Lyapunov exponents for these orbits}.
This is not to say that the orbits will definitely be chaotic, just that 
there will be instability and a positive Lyapunov exponent.

\begin{figure}[t]
\vspace{48mm} 
\includegraphics{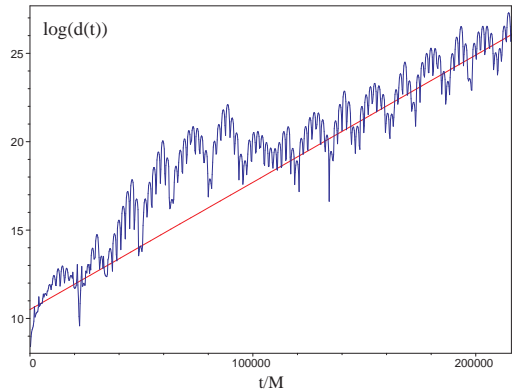}
\vspace{5mm}
\caption{Determining the Lyapunov exponent for an orbit taken from Fig.\ 3
of Ref.\ [3].
\label{lefbb}}  
\end{figure}

Chaos was detected in the Post-Newtonian 
(PN) expansion of the two-body problem 
for rapidly spinning bodies \cite{mepn} using the method of  
fractal basin boundaries.
One fully expects positive Lyapunov exponents at least very
near the fractal basin 
boundaries, if not elsewhere in the phase space.  
We sampled some orbits near the boundary and find positive Lyapunov
exponents as illustrated
by the positive slope in Fig.\ \ref{lefbb}. 
The reason for the discrepancy seems to be that the authors of Ref.\ \cite{sr}
define the maximum exponent as ``the Cartesian distance between the
dimensionless 12-component coordinate vectors
[${\bf \vec r, \dot{ \vec r}, \vec S_1, \vec S_2}$]
and [${\bf {\vec r}^\prime, \dot{ \vec r^\prime}, \vec S^\prime_1, \vec S^\prime_2}$]
of two nearby trajectories'' \cite{sr}. 
However, {\it this is not a Lyapunov exponent}.
The Lyapunov exponent is obtained by linearizing the equations of motion
about a given trajectory.
An {\it approximation} to the Lyapunov exponent can be made using the
Cartesian distance between two trajectories by continually
rescaling the shadow trajectory so that it is always infinitesimally 
close to the original trajectory as was done by
\cite{maeda}. However, the result does depend on the rescaling
and can give false answers.
A more thorough treatment 
of the Lyapunov exponents and their interpretation 
is given elsewhere \cite{us}.

\

Chaos has not been ruled out rather it has been confirmed by 
positive Lyapunov exponents.  In fact, 
an even stronger claim can be made.  The two-body problem has not been 
solved in relativity and it would not be too daring to 
conjecture that it will never be solved.  The system shows every indication
of being fully non-integrable.  

Regardless of how 
we'd like nature to behave to make our lives as observers
easier, she may not be so accommodating.
Understandably there is a sense of urgency to the removal of any obstacles
to the detection of gravitational waves.  However, chaos need not be
a terrible obstacle.
There are ways that chaos can
enhance signals or amplify salient features in the sky \cite{me} to aid
our quest to observe gravitational
waves.

\end{document}